\documentclass{emulateapj}

\usepackage{epsfig}
\usepackage{amsmath}
\usepackage{amssymb}
\usepackage{natbib}
\usepackage{graphicx}
\usepackage{ulem}

\shorttitle{Magnetic fields in the Antennae}
\shortauthors{Kotarba et al.}

\begin{document}

\title{Simulating magnetic fields in the Antennae galaxies}
\author{H. Kotarba$^{1,2}$, S. J. Karl$^{1}$, T. Naab$^{1,3}$, P. H. Johansson$^{1}$, K. Dolag$^{3}$, H. Lesch$^{1}$ \& F. A. Stasyszyn$^{3}$}
\email{kotarba@usm.lmu.de}

\affil{$^1$ University Observatory Munich, Scheinerstr.1, D-81679 Munich, Germany}
\affil{$^2$ Max Planck Institute for Extraterrestrial Physics, Giessenbachstrasse, D-85748 Garching, Germany}
\affil{$^3$ Max Planck Institute for Astrophysics, Karl-Schwarzschild-Str. 1, D-85741 Garching, Germany}

\begin{abstract}
We present self-consistent high-resolution simulations of NGC4038/4039 (the ''Antennae galaxies``) including star formation, supernova feedback and magnetic fields performed with the \textsl{N}-body/SPH code $\textsc{Gadget}$, in which magnetohydrodynamics are followed with the SPH method. We vary the initial magnetic field in the progenitor disks from $10^{-9}$ to $10^{-4}$ G. At the time of the best match with the central region of the Antennae system the magnetic field
has been amplified  by compression and shear flows to an equilibrium field value of $\approx 10$ $\mu$G, independent of the initial seed field. These simulations are a proof of the principle that galaxy mergers are efficient drivers for the cosmic evolution of magnetic fields. We present a detailed analysis of the magnetic field structure in the central overlap region. Simulated radio and polarization maps are in good morphological and quantitative agreement with the observations. In particular, the two cores with the highest synchrotron intensity and ridges of regular magnetic fields between the cores and at the root of the southern tidal arm develop naturally in our simulations. This indicates that the simulations are capable of realistically following the evolution of the magnetic fields in a highly non-linear environment. We also discuss the relevance of the amplification effect for present day magnetic fields in the context of hierarchical structure formation.
\end{abstract}


\label{firstpage}
\section{Introduction}

Within the framework of the hierarchical galaxy formation picture, galaxies assemble and evolve via mergers of smaller progenitor galaxies (e.g. \citealp{White&Rees1978}, \citealp{White&Frenk1991}). Thus, galaxy interactions are essential for the understanding of structure formation. In the bottom-up-picture of structure formation dwarf galaxies merge to form the first galaxies at an early epoch of the universe. Later, there is still a continuous merging of fully evolved galaxies. The further growth of galaxies progresses through a combination of mergers and diffuse accretion of gas.

Interactions of galaxies change their dynamics drastically (see e.g. \citealp{Toomre&Toomre1972}, \citealp{Barnes1992}, \citealp{Hernquist1994}, \citealp{Barnes1999} and \citealp{Burkert&Naab2003}, \citealp{Gonzalez-Gracia2006}) as the gravitational potential is changing rapidly during the interaction.  Since the gas component is dissipative and most sensitive to changes of the gravitational potential, it is strongly affected during the interaction and driven to the galaxy centers, eventually causing bursts of star formation (\citealp{Barnes&Hernquist1992}, \citealp{Mihos&Hernquist1994}, \citealp{Barnes&Hernquist1996}, \citealp{Bekki&Shioya1998}, \citealp{Springel2000}, \citealp{Barnes2002}, \citealp{Bournaud2005}, \citealp{Cox2006}, \citealp{Naab&Jesseit2006}, \citealp{Robertson2006}, \citealp{Cox2008}, \citealp{Hopkins2008}). So far simulations of interactions and mergers of disk galaxies have only been investigated with respect to changes in stellar dynamics, gas flows, star formation (SF) or formation of central supermassive black holes (\citealp{DiMatteo2005Natur}, \citealp{Springel2005}, \citealp{Springel2005BlackHoles}, \citealp{Robertson2006}, \citealp{DeBuhr2009}, \citealp{Johansson2009}, \citealp{Johansson2009_2}). However, the dramatic impact of mergers on the gas flows will directly affect the magnetic fields of the systems (and vice versa) via the induction equation of magnetohydrodynamics (MHD) and the Lorentz force. The magnetic fields will change their morphology following the motion of the gas and will be amplified by shocks and gas inflow.

Changes in the magnetic field structure, on the other hand, might influence gas flows, local collapse and the morphology as well as the star formation activity. Local, interacting galaxies are the perfect laboratories for investigating the effects associated with their magnetic fields. However, the timescales for galaxy mergers are far too long to observe these processes directly. The only observational possibility to study the time evolution of mergers is to consider different systems at different evolutionary stages. However, the available sample of interacting nearby galaxies is too small to investigate the evolution of magnetic fields in detail. Thus, numerical simulations pose a promising tool to study the magnetic field evolution in interacting systems.

The structure of an interacting system strongly depends on the properties of the progenitor galaxies. Thus, matching observed nearby interacting systems with simulations in space and time can give us an idea of the properties of their progenitors, e.g. their sizes, gas fractions and relative velocities. Furthermore, comparing simulated systems with observations helps to asses the performance of the applied numerical method. Numerical methods supported by these comparisons can then be used to study processes in the early universe, when galaxies were very different from present-day galaxies. High resolution simulations of the formation of individual galaxies in a full cosmological context (see e.g. \citealp{Naab2007}) including magnetic fields could help us in understanding the processes leading to the magnetization of the universe. This type of study would complement earlier semi-analytical studies that investigated the possibility of magnetic field seeding by galactic winds in a cosmological context (\citealp{Bertone2006}).

The standard theory of magnetic field amplification in galaxies is the so called Galactic Dynamo based on the mean field theory (see \citealp{Kulsrud1999} for a review). Within this theory, turbulent motions of the ionized gas driven by stellar activity lead to the generation of a random magnetic field ($\alpha$-effect). This random magnetic field (particularly its radial component) can then be amplified by the differential rotation of the galaxy ($\Omega$-effect), leading to an efficient dynamo action which results in an exponential growth of the magnetic field (the $\alpha\Omega$-dynamo). However, dynamos may probably only work efficiently if magnetic helicity is transported away from the differentially rotating disc (\citealp{Brandenburg2005}). \citet{Gressel2008} performed high-resolution box-simulations which demonstrate that a dynamo may operate if supernova explosions release magnetic helicity from the disc. However, for an efficient magnetic helicity transport out from a galactic disk, galactic winds or galactic fountains may be required. This might be a problem particularly for massive galaxies due to the deeper potential well. The fact that it is difficult to get an efficient dynamo is generally addressed as the dynamo problem. Different solutions, e.g. turbulence driven by large-scale SN-bubbles (\citealp{Ferriere1992}) or the Cosmic Ray Dynamo (\citealp{Hanasz2009}) have been proposed. These solutions describe the exponential growth of a small-scale magnetic seed field which is amplified up to present-day values within several Gyr. However, recent observations indicate that magnetic fields in galaxies have been already very strong (comparable to present-day galactic magnetic fields) at very high redshifts, at a time when the universe was only $t \approx$ 6 Gyr old ($z\approx 1$) (\citealp{Bernet2008}). Former observations of damped Ly-$\alpha$ systems by \citet{Wolfe1992} indicate that progenitors of galactic discs had magnetic fields of a few $\mu$G even at $z\approx2$ ($t\approx 3$ Gyr). The very fast amplification required to generate the strong magnetic fields at high redshifts can probably not be achieved with any Galactic Dynamo model (see e.g. \citealp{Arshakian2009}). Thus, alternative possibilities for the amplification of galactic magnetic fields on shorter timescales need to be explored. \citet{Lesch&Chiba1995} have shown analytically that strong magnetic fields in high redshift objects can be explained by the combined action of an evolving protogalactic fluctuation and electrodynamic processes providing magnetic seed fields. \citet{Wang&Abel2009} performed numerical simulations of the formation of disc galaxies within an collapsing halo imposing a uniform initial magnetic field of $10^{-9}$ G. The initial field was amplified by three orders of magnitude within approximately 500 Myr of evolution. The amplification might be due to the combined effects of magnetic field compression during the collapse and amplification of the uniform initial field by differential rotation as studied also in \citet{me2009}. These studies indicate, that the amplification of magnetic fields might be a natural part of the galaxy formation process. However, interactions of galaxies, which were more frequent at earlier times, pose another promising possibility.

Although it would be worthwile to consider cosmological studies of structure formation including magnetic fields in the long run, numerical studies of interacting magnetized systems in the local universe may serve as a first step towards a more complete scenario. These studies help us in understanding the morphological evolution of galaxies, their star formation histories (\citealp{Springel2005Natur}, \citealp{Cox2008}, \citealp{Bournaud2007}, \citealp{DiMatteo2008}, \citealp{Jesseit2009}, \citealp{Naab&Ostriker2009}), and as we will show in this paper also their magnetic histories. The system NGC 4038/39, also known as the Antennae galaxies, is one of the best examples for an ongoing local merger.  It is also the by far best observed interacting galaxy system.

In this paper we present further steps towards a more complete numerical representation of the Antennae system as a prototype for interacting galaxies. For the first time we will follow the evolution of the magnetic field in a galaxy interaction simulation. We also address the general question whether smoothed particle magnetohydrodynamics (SPMHD) is capable of following the evolution of magnetic fields in interacting systems at reasonable accuracy. In a previous paper we have shown that SPMHD is well suited for following the evolution of magnetic fields in isolated disk galaxies (\citealp{me2009}) so the study of interacting systems is a natural extension of this earlier study.

The paper is organized as follows: Section \ref{ANT_PROP} summarizes the properties of the Antennae system as known from observations and theory. In section \ref{SIMULATIONS} we describe our numerical methods (section \ref{NUMERICS}), the setup of the isolated disks (section \ref{SETUP_ISOLATED}) and the match to the observed Antennae system (section \ref{SETUP_MATCH}). A detailed analysis of the evolution of the system is presented in section \ref{EVOUTION_OF_ANTENNAE}, where we discuss the evolution of the magnetic field (section \ref{EVOUTION_OF_ANTENNAE_MAG}) the numerical stability of our simulations (section \ref{NUMERICAL_STABILITY}) and the self-regulation of the magnetic field amplification (section \ref{PRESSURES}). In section \ref{RADIO_MAPS} we describe our method to calculate artificial radio maps (section \ref{RADIO_METHOD}) and present applications to the isolated disk and the Antennae simulations (section \ref{RADIO_APPLICATIONS}). The artificial radio maps derived from the simulations can be compared against the radio observations of the system, thus providing a further tool for constraining the numerical model and method. We conclude in section \ref{SUMMARY} and briefly discuss the relevance of our simulations in the context of hierarchical structure formation.

\section{Properties of the Antennae systems}\label{ANT_PROP}

The Antennae system is relatively nearby, the estimated distances range from 13 to 25 Mpc. The smaller distances are favored by methods based on photometry of the red giant branch (\citealp{Saviane2008}), whereas the larger distances are estimated from observations of the Type Ia supernova 2007sr in the southern tail (\citealp{Schweizer2008}). But note also that sometimes even values up to $d = 29$ Mpc have been adopted in the literature (\citealp{FabbianoEtAl2001}, \citealp{ZezasFabbiano2002}). In this paper, we apply the conventional distance of 22 Mpc for all relevant flux calculations. Given the large variety of high quality observations (\citealp{Withmore1999}, \citealp{Neff2000}, \citealp{Wilson2000}, \citealp{Hibbard2001}, \citealp{Chyzy&Beck2004}, \citealp{Wang2004}, \citealp{Brandl2005}, \citealp{Zezas2006}, \citealp{Brandl2009}) several authors tried to find initial conditions for simulations representing the Antennae system. \citet{Toomre&Toomre1972} first presented restricted three-body simulations which already explained the formation of tidal arms and bridges as a result of tidal interaction during the merger. Follow-up investigations confirmed the early results by studying the detailed galactic dynamics using self-consistent, multiple-component galaxy models \citep{Barnes88}. Further studies added star-formation \citep{Mihos+93} to the modeling process and aimed at constraining the influence of dark halo mass profiles on the development and morphology of the tidal tails \citep{Dubinski+96}.

Recently, \citet{Karl2010} developed a new model of the system, not only focussing on its plane-of-sky appearance, but also on fitting the observed line-of-sight velocity structure (see also \citealp{Karl2008AN}). This study, alongside with new observations \citep{ZhangGaoKong2009}, suggests that the localized intense starburst sites observed in the overlap-region
can be explained as the imprint of the interpenetrating process of the two merging disks following their second encounter. These results contrast with previous numerical simulations which find the current orbital phase of the Antennae system to lie somewhere between the first and the second closest encounters (e.g. \citealp{Toomre&Toomre1972}, \citealp{Barnes88}, \citealp{Dubinski+96}).

\begin{figure}
\begin{center}
\epsscale{.90}
\plotone{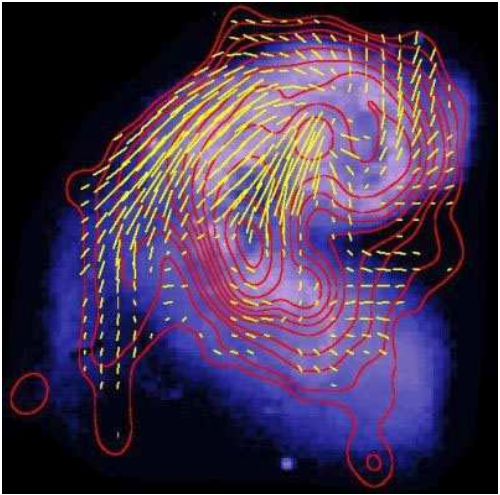}
  \caption{Total synchrotron emission (contours) and magnetic field vectors of polarized intensity at 4.86 GHz based on VLA data (yellow), overlaid on a DSS image (blue - white background) (Digitized Sky Survey, Palomar and UK Schmidt telescopes). The contour levels are 0.005, 0.12, 0.30, 0.53, 1.2, 2.1, 3.3, 5.3, 9.0, 17 and 24 mJy/beam-area. The resolution is 17''$\times$14''. Credit: \citet{Chyzy2005}}
  {\label{Antennae2}}
\end{center}
\end{figure}

In this paper we use the model of \citet{Karl2010} and focus on the central region of the Antennae system and its magnetic fields.
Fig. \ref{Antennae2} shows a DSS image (Digitized Sky Survey (Blue), conducted with the Palomar and UK Schmidt telescopes and digitized by the Catalogs and Surveys Group of the Space Telescope Science Institute) of the central region of the Antennae system, i.e. the galactic disks and bases of the tidal tails. Overlaid are contours of total radio synchrotron emission (tracing the total magnetic field energy) and magnetic field vectors (derived from polarized intensity). The strength of the magnetic field is $20$ $\mu$G on average. The highest values of more than 30 $\mu$G are reached in the overlapping region and the centers of the galaxies (\citealp{Chyzy&Beck2004}). Thus, the magnetic field is roughly twice as strong as the typical values of 5 to 10 $\mu$G observed in isolated spiral galaxies (e.g. \citealp{Beck1996}, \citealp{Beck2007}, \citealp{Krause2009}). On the upper left (east), there is a large region with highly ordered magnetic field lines, most probably tracing the gas flow at the root of the lower (southern) tidal tail. This gas is also visible as a HI ridge which extends far out along the southern tail (\citealp{Hibbard2001}). The structure of the magnetic field most likely traces the recent gas motions induced by tidal forces during the merger. Apparently, not much of the magnetic field structure of the progenitor galaxies have survived the interaction. The progenitors were presumably typical spirals with a spiral magnetic field pattern (see e.g. \citealp{Beck2009}). As the magnetic field is tightly linked to the motion of the gas, the structure of the field in a system which has recently undergone a violent interaction should mainly resemble the recent kinematic evolution. It does not depend on long-term processes like the Galactic Dynamo, which is believed to be important in isolated spiral galaxies (e.g. \citealp{Brandenburg2005}, \citealp{Gressel2008}, \citealp{Beck2009}, \citealp{Dubois&Teyssier2009}, \citealp{Elstner2009}, \citealp{Gissinger2009}, \citealp{Hanasz2009}). In other words, nonlinear systems lose the memory of their initial conditions. Hence, numerical studies of the kinematics of merging systems including magnetic fields should be able to represent the observed magnetic fields in nearby interacting systems also without including long-term processes.

\section{Simulations}\label{SIMULATIONS}

\subsection{Numerical methods}\label{NUMERICS}
All simulations were performed with the $\mathrm{N}$-body/SPH-code $\textsc{Gadget}$ (\citealp{SpringelGadget}). Gravitational interactions between the particles are evaluated with a hierarchical tree method (\citealp{Barnes&Hut1986}). The dynamics of Lagrangian fluid elements are followed using a SPH formulation which conserves both energy and entropy (\citealp{Springel&Hernquist2002}) including the evolution of magnetic fields
which was implemented and tested by \citet{GadgetMHD}. The code has already been used to investigate the evolution of magnetic fields in isolated spiral galaxies (\citealp{me2009}) and to compare different implementations of the SPH formulations and implementations in the SPH Code VINE (\citealp{VINEI}, \citealp{VINEII}). These studies demonstrated the importance of a sensible treatment of the numerical divergence of the magnetic field ($\nabla\cdot\textbf{B}$) in SPH simulations, as it can lead to artificial magnetic field growth. An implementation utilizing Euler potentials, which by construction poses a $\nabla\cdot\textbf{B}$-free prescription of magnetic fields (see \citet{Price&Bate2007} for more details) circumvents this problem. However, using the Euler potentials, the magnetic field is essentially mapped on the initial particle arrangement. Thus, if the initial configuration significantly changes shape during the simulation, regions carrying conflicting values of Euler potentials (i.e. values, which do no longer allow for a finite and unambiguous calculation of their gradients) can overlap and the ability of the Euler potentials to represent the magnetic field correctly is lost. This can lead to defective magnetic field calculations, especially in kinematically vigorous systems like interacting galaxies (see also \citealp{Brandenburg2009}). Therefore, all simulations presented in this paper have been performed using the standard (direct) magnetic field implementation. In contrast to \citet{me2009} we now also apply the Lorentz force and artificial magnetic dissipation applying an artificial magnetic dissipation constant of $\alpha_B=0.5$. The latter does not only allow for magnetic field redistribution and reconnection, but also lowers the numerical divergence as it helps to smooth artificially high magnetic fields arising from intrinsic constraints of the numerical prescription. In this sense, it poses a regularization scheme similar to smoothing of the magnetic field. Both schemes reduce the numerical noise and $\nabla\cdot\textbf{B}$ errors by a similar amount (\citealp{GadgetMHD}). However, the dissipation scheme is based on physical considerations, whereas the smoothing scheme is completely artificial. Thus, using the dissipation scheme, the factor $h\nabla\cdot\textbf{B}/|\textbf{B}|$ (where $h$ is the so called smoothing length, which poses the typical length scale of spatial derivatives in SPH calculations) is restricted to a value of approximately unity. Values of order unity have been shown to be low enough to guarantee a physically meaningful evolution of the magnetic fields in SPH simulations, particularly preventing artificial magnetic field growth. This threshold is actually defined by simulations using Euler potentials, for which the numerical divergence measure $h\nabla\cdot\textbf{B}/|\textbf{B}|$ is of order unity although the physical divergence is zero by definition (see \citealp{me2009} and section \ref{NUMERICAL_STABILITY} for more details).

Furthermore, we do not use a viscosity limiter as suggested by \citet{Balsara1998}, because applying this limiter resulted in an increased growth of the magnetic field. This is most likely a numerical artefact, as the limiter lowers the viscosity in regions of strong shear flows, thus suppressing velocity diffusion and leading to a higher velocity dispersion and higher velocity gradients, which in turn lead to artificially enhanced magnetic field growth (\citealp{me2009}).

All simulations are performed including radiative cooling assuming a primordial gas composition together with a homogeneous and time-independent extragalactic UV background (\citealp{Haardt&Madau1996}). We include star formation and the associated supernova feedback, but exclude explicit supernova-driven galactic winds, following the sub-resolution multiphase model developed by \citet{Springel&Hernquist2003}, in which the ISM is treated as a two-phase medium (\citealp{McKee1977}, \citealp{Johansson&Efstathiou2006}): Cold clouds are embedded in a tenous hot gas at pressure equilibrium. Stars form from the cold clouds in regions were $n>n_\mathrm{th}=0.128$ cm$^{-3}$ with the shortlived stars supplying an energy of $10^{51}$ ergs to the surrounding gas by supernovae. The threshold density, $n_\mathrm{th}$, is determined self-consistently in the model by requiring that the equation-of-state (EOS) is continuous at the onset of star formation. The parameters governing the model (see Tab. \ref{tab1}) are set to produce a star formation rate of $\sim 1 M_{\odot}$ yr$^{-1}$ for a Milky Way-like galaxy in isolation.

The implementation used in this paper has been tested in detail (\citealp{SpringelGadget1}, \citealp{SpringelGadget}, \citealp{Springel2005}, \citealp{GadgetMHD}) and fulfills the established requirements for numerical methods. Particularly, \citet{GadgetMHD} have shown that the MHD-implementation performs well in various test problems, including different shock tube problems, the Fast Rotator (\citealp{Balsara&Spicer1999}), the Strong Blast (e.g. \citealp{Balsara&Spicer1999}) and the Orszag-Tang Vortex (\citealp{Orszag&Tang1979}).

\subsection{Setup}\label{SETUP}
\subsubsection{Isolated disks}\label{SETUP_ISOLATED}

The Antennae system has most likely formed through the interaction of two formerly isolated spiral galaxies. In this section we present the properties of the isolated progenitor model disks used in our simulations. The initial conditions for the spiral galaxies are realized using the method described by \citet{Springel2005} which is based on \citet{Hernquist1993}. The galaxies consist of a \citet{Hernquist1990} profile cold dark matter halo, a rotationally supported exponential stellar disk, an exponential gas disk and a stellar \citet{Hernquist1990} bulge component (see \citealp{Karl2010} for more details). The halo, stellar disk and bulge particles are collisionless $\mathrm{N}$-body particles. The gas is represented by SPH particles.

Possible initial conditions resulting in a good fit of the present-day properties of the Antennae galaxies have been tested
in a large parameter study by \citet{Karl2010} (see also section \ref{SETUP_MATCH}). In this paper we present results using the initial condition parameters of this study which result in the best match to the central region of the Antennae system.
The parameters describing the initial conditions of the two galaxies can be found in Table \ref{tab1}. Particle numbers and softening lengths are listed in Table \ref{tab2}. The disks are kinematically stable if evolved in isolation as has been shown in detail in \citet{me2009}. In the following we thus only address the evolution of the magnetic fields.

\begin{deluxetable}{lll}
\tabletypesize{\scriptsize}
\tablecaption{Parameters of initial setup\label{tab1}}
\tablewidth{0pt}
\startdata\tableline\tableline
\multicolumn{3}{c}{\textsc{Disk parameters}}\\\tableline
  total mass & $M_\mathrm{tot}$   &  $1.34\times 10^{12}M_\odot$    \\
  disk mass  & $M_\mathrm{disk}$  &  0.075 $M_\mathrm{tot}$   \\
  bulge mass & $M_\mathrm{bulge}$ &  0.025 $M_\mathrm{tot}$ \\
  mass of the gas disk & $M_\mathrm{gas}$ & 0.2 $M_\mathrm{disk}$  \\
  exponential disk scale length & $l_D$   & 8.44 kpc \\
  scale height of the disk      & $h_D$     & 0.2 $l_D$ \\
  bulge scale length            & $l_B$   & 0.2 $l_D$ \\
  spin parameter              & $\lambda$ & 0.1 \\
  virial velocity of the halo & $v_\mathrm{vir}$  & 160 km s$^{-1}$ \\
  half mass radius            & $R_\mathrm{half}$ & $\approx$12 kpc\\
  half mass circular velocity & $v_\mathrm{half}$ & $\approx$249 km s$^{-1}$ \\
  half mass rotation period   & $T_\mathrm{half}$ & $\approx$295 Myr \\
  initial magnetic field      & $B_0$ & $10^{-9}$ to $10^{-6}$ G \\ \tableline
  \multicolumn{3}{c}{\textsc{Multi-Phase model parameters}}\\\tableline
  gas consumption timescale   & $t_\mathrm{MP}$ & 8.4 Gyr \\
  mass fraction of massive stars & $\beta_\mathrm{MP}$ & 0.1 \\
  evaporation parameter & $A_0$ & 4000 \\
  effective SN temperature & $T_\mathrm{SN}$ & $4\times10^8$ K \\
  cold cloud temperature & $T_\mathrm{CC}$ & 1000 K
\enddata
\end{deluxetable}

\begin{deluxetable}{lcc}
\tabletypesize{\scriptsize}
\tablecaption{Particle numbers and softening lengths\label{tab2}}
\tablewidth{0pt}
\tablehead{\colhead{Component} & \colhead{initial particle number} & \colhead{fixed gravitational}\\
 & & \colhead{softening length $\epsilon$ [pc]$^{\rm{a,b}}$}}
\startdata
  Halo      &  $4.0\times10^5$     & 80/$h$  \\
  Disk      &  $4.8\times10^5$   & 20/$h$  \\
  Bulge     &  $2.0\times10^5$   & 20/$h$  \\
  Gas       &  $1.2\times10^5$   & 20/$h$  \\
  Stars     &  $0$              & 20/$h$  \\
  Total     &  $1.2\times10^6$  & -
\enddata
\tablenotetext{a}{The minimum SPH smoothing length for the gas particles is 1.0$\epsilon$.}
\tablenotetext{b}{The Hubble constant is assumed to be $h=0.71$ in this paper.}
\end{deluxetable}

\begin{figure}
\begin{center}
\epsscale{1.1}
\plotone{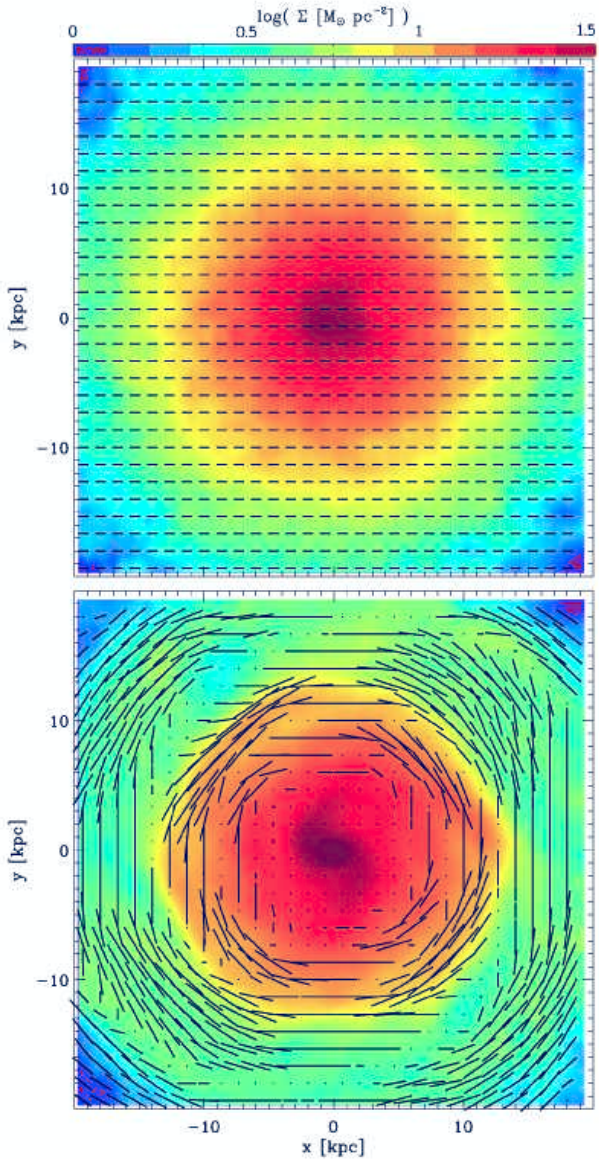}
  \caption{Gas surface density $\Sigma$ at time $t=0$ Myr (upper panel) and $t=400$ Myr (lower panel), overlaid with magnetic field vectors for the simulation with $B_0=10^{-6}$ G. The length $l$ of the vectors is normalized to a minimal value $B_\mathrm{min}=B_0/\sqrt{2}$ and displayed logarithmically according to $l=5\log{(B/B_\mathrm{min})}$, i.e. $l = 0$ corresponds to $B \approx B_\mathrm{min}$ or smaller and $l = 5$ to $B = 10 B_\mathrm{min}$.
  \label{bfield}}
\end{center}
\end{figure}

For simplicity, the initial seed magnetic field is assumed to be homogeneous with only one non-vanishing component of $B_x=B_0$. This choice is justified, as it takes more than one Gyr of dynamic evolution until the present plane-of-sky-appearance of the system has developed in our simulations. Thus, the particular structure of the initial magnetic field should not be of significance for the final result. We use two different values for the initial field, $B_0=10^{-9}$ G and $B_0=10^{-6}$ G for the isolated galaxies, and additionally two intermediate values, $B_0=10^{-8}$ G and $B_0=10^{-7}$ G for the Antennae simulations. The smallest value of $B_0=10^{-9}$ G is the typical value of the observed intergalactic magnetic field $B_\mathrm{IGM}$ (see e.g. \citealp{Kronberg2008}) and the highest, $B_0=10^{-6}$ G, is motivated by the typical value of several $\mu$G observed in spiral galaxies. As much larger or much smaller values are not observed, these values cover the range of realistic initial fields. However, we have also performed a simulation of the Antennae system with an initial magnetic field value of $B_0=10^{-4}$ G in order to study the physical behaviour of the system in an extreme situation. We do not include neither large-scale dynamo processes, nor turbulent motions on scales smaller than $\approx$ 100 pc which are not resolved in our simulations. The mean velocity dispersion $\sigma=\sqrt{\langle \vec{v}^2\rangle_\mathrm{n.n.} - \langle \vec{v}\rangle_\mathrm{n.n.}^2}$ (where the mean is taken over the nearest $64\pm5$ neighbors within the smoothing kernel) during the isolated disc simulation is of order of 5 km s$^{-1}$ with approximately 30\% of the particles having dispersions $>5$ km s$^{-1}$ and only a few percent $>10$ km s$^{-1}$. These values are somewhat lower than the values found in recent grid simulations by \citet{Wang&Abel2009} (see Fig. 3 of their paper) and \citet{Agertz2009}. They both find typical dispersion values of approximately 10 km s$^{-1}$ in their comparable disc galaxy simulations. However, these authors use lower temperature floors for the dense gas component found in the star-forming regions, resulting in a clumpier disc structure and thus probably in an enhanced turbulence in the hot diffuse component of their discs. This may explain the discrepancy in the measured velocity dispersions. Since the dispersion values in our simulations are rather low, we do not expect any significant amplification of the magnetic field in the isolated galaxies. Consequently, the magnetic field gets only redistributed during the simulation, developing a spiral pattern as the differential rotation continues to wind it up (Fig. \ref{bfield}), while the overall value of $|\vec{B}|$ remains of order $B_0$ throughout the simulation. After the magnetic field has been wound up by differential rotation, it is highly ordered in the disc region ($r>5$ kpc) and more ''turbulent`` in the inner region of the galaxy. Thus, the inner magnetic field is not visible in the lower panel of Fig. \ref{bfield} due to the averaging calculation of our plotting routines.

\begin{figure}
\begin{center}
\epsscale{1.2}
\plotone{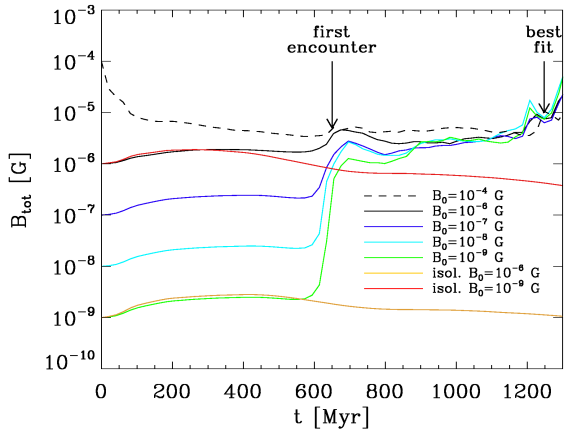}
  \caption{$B_{tot}=\sqrt{B_x^2+B_y^2+B_z^2}$ as a function of time for the Antennae simulations with an initial field of $B_0=10^{-9}$ G (green line), $B_0=10^{-8}$ G (blue line), $B_0=10^{-7}$ G (dark blue line), $B_0=10^{-6}$ G (black line) and $B_0=10^{-4}$ G (black dashed line), respectively, and for the progenitor disks simulations with $B_0=10^{-9}$ G (orange line) and $B_0=10^{-6}$ G (red line), respectively. The magnetic field of the isolated disks does not evolve significantly. For the mergers the field is amplified to $\approx 10 \mu$G independent of the initial field strength in the disks.
  \label{Bevol}}
\end{center}
\end{figure}

The evolution of the absolute value of the magnetic field as a function of time is shown in Fig. \ref{Bevol} for $B_0=10^{-6}$ G (red line) and $B_0=10^{-9}$ G (orange line). In the beginning of the simulation, the initially homogeneous magnetic field gets wound up and thereby amplified due to the differential rotation and associated shear flows by roughly a factor of two (see also \citealp{me2009}). After approximately one half mass rotation period the magnetic field has been redistributed to a mostly toroidal pattern by the winding process, and the amplification ceases. In the subsequent evolution, the strength of the field decreases slowly due to magnetic dissipation, which is the only process causing magnetic field diminution in our simulations. The velocity dispersion of the gas also leads to the development of a $z$-component of the magnetic field (not shown) which, however, remains smaller than all other components by more than one order of magnitude throughout the simulation. In summary, the magnetic field gets redistributed to form a spiral pattern (Fig \ref{bfield}) and retains on average its initial value throughout one Gyr of evolution. This behaviour is the same for both the weaker and the stronger initial magnetic field.

\begin{figure}
\begin{center}
\epsscale{1.2}
\plotone{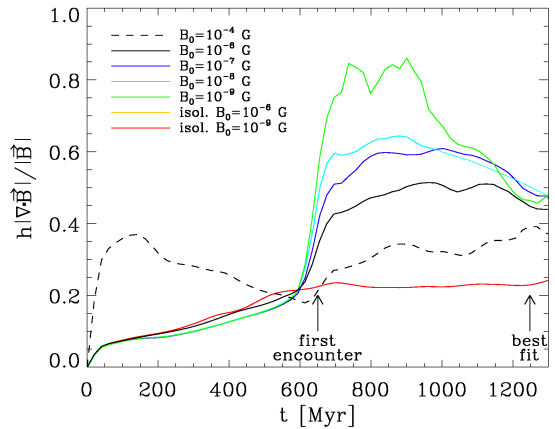}
  \caption{$h\nabla\cdot\textbf{B}/|\textbf{B}|$ as a function of time for the Antennae simulations with an initial field of $B_0=10^{-9}$ G (green line), $B_0=10^{-8}$ G (blue line), $B_0=10^{-7}$ G (dark blue line), $B_0=10^{-6}$ G (black line) and $B_0=10^{-4}$ G (black dashed line), respectively, and mean divergence for isolated simulations with $B_0=10^{-6}$ G (red line). The values stay below the tolerance value of unity throughout the simulation in every run.
  \label{divB}}
\end{center}
\end{figure}

Fig. \ref{divB} shows the mean numerical divergence $h\nabla\cdot\textbf{B}/|\textbf{B}|$ as a function of time in isolated galaxy simulations (red line) with $B_0=10^{-6}$ G. The mean was taken over three simulations with the magnetic field in the plane of the disk and inclined as in the setup of the Antennae simulation (see section \ref{SETUP_MATCH}), respectively. Although the numerical noise increases with time, it remains clearly below the tolerance value of unity (see also section \ref{NUMERICAL_STABILITY}).

The SFR in the isolated disks is roughly constant throughout the simulations (not shown). Starting at a value of approximately 2 $M_\odot$ yr$^{-1}$ and then decreasing slightly to approximately 1.7 $M_\odot$ yr$^{-1}$ after 1.3 Gyr of evolution due to gas consumption. There is no significant difference in the evolution of the SFR compared to the same simulation without any magnetic field, indicating that the presence of a global magnetic field of order $10^{-6}$ G or lower does not affect the gas flow enough to hinder or abet the collapse of gas.

\subsubsection{The match to the Antennae system}\label{SETUP_MATCH}

The simulations presented here are taken out of a suit of self-consistent simulations designed as a large parameter study to
match the morphological and kinematical properties of the Antennae (see \citealp{Karl2010}). In this study, we initially set two equal-mass galaxies, each modeled as in section \ref{SETUP_ISOLATED} and residing in its own dark matter halo, on nearly-parabolic Keplerian two-body orbits with given ellipticity $e$, pericenter distance $r_{\rm{p}}$, and initial separation $r_{\rm{sep}}$. The disk orientation in the orbital plane is given by a pair of angles ($\iota$, $\omega$), which, for each galaxy, specify the adopted inclination with respect to the orbital plane and
the pericentric argument \citep{Toomre&Toomre1972}. There is no hot gas component surrounding the galaxies initially. The initial field is assumed to be homogeneous with only one non-vanishing component of $B_x=B_0$. After the simulation has finished we determine the time of best match, the
viewing direction of the observer, a common center-of-mass, and a distance scale factor $\mathcal{L}$ in order to create a mock observation which can be compared to projections of the HI data cube from \citet{Hibbard2001}. If the result does not prove satisfactory up to a level admissible by optical inspection, the simulation is repeated choosing a different set of initial parameters. Several key parameters regarding the elliptical orbit, the relative orientation of the galaxy disks, and the internal structure of the progenitor galaxies are varied in order to find the best match (for details, see \citealp{Karl2010}). The final parameters used in this study are shown in Tab. \ref{Tab:OrbitAnalysisParameters}.

Starting on their initially set orbit, both galaxies evolve corresponding to their isolated evolution (section \ref{SETUP_ISOLATED}) until they reach the point of their first closest approach ($t \approx 650$ Myr). At this time, the prominent tidal arms, which we use as tracers for the dynamical history of the encounter, start to develop. On the other hand, the detailed structure of the galactic main bodies can only be seen in our simulations resulting from the recent splash during the second encounter ($t \approx 1180$ Myr). The time of best fit, i.e. the time, at which the simulation matches
the appearance of the Antennae system in the sky and the observed line-of-sight velocities, is reached at $t_\mathrm{BM} \approx 1250$ Myr.

\begin{deluxetable}{lll}
\tabletypesize{\scriptsize}
\tablecaption{Antennae simulation parameters\label{Tab:OrbitAnalysisParameters}}
\tablewidth{0pt}
\startdata
\tableline\tableline
\multicolumn{3}{c}{\textsc{Initial orbit parameters}}\\
\tableline
disk orientation & NGC 4038 & NGC 4039\\
     $\iota$ & $60^{\circ}$ & $60^{\circ}$\\
     $\omega$ & $30^{\circ}$ & $60^{\circ}$\\
\tableline
ellipticity & $e$ & $0.96$ \\
pericenter distance & $r_{\rm{p}}$ & $7$ kpc$\thinspace$h$^{-1}$\\
initial separation &$r_{\rm{sep}} = r_{\rm{vir}}$ & $160$ kpc$\thinspace$h$^{-1}$\\
\tableline
\multicolumn{3}{c}{\textsc{Analysis parameters}}\\
\tableline
time of best match & $t_{\rm{BM}}$ & $1.25$ Gyr\\
direction to observer$^{\rm{a}}$ & ($\theta$,$\psi$,$\phi$)&  (208,282,0)\\
distance scale & $\mathcal{L}$ & $2.0$
\enddata
\tablenotetext{a}{The viewing direction is specified by a series of rotations in the following order around the $x$-, $y$-, and $z$-axis.}
\end{deluxetable}

\subsection{Evolution of the Antennae system}\label{EVOUTION_OF_ANTENNAE}
\subsubsection{Magnetic field evolution}\label{EVOUTION_OF_ANTENNAE_MAG}

We have run several simulations using the setup described in section \ref{SETUP_MATCH}. The initial field was again assumed to be homogeneous with only one non-vanishing component of $B_x=B_0$ at the beginning of the simulation. We performed simulations with five different initial magnetic field strengths of $B_0=10^{-9}$, $10^{-8}$, $10^{-7}$ $10^{-6}$ and $10^{-4}$ G for comparison.

\begin{figure*}
\begin{center}
\epsscale{0.9}
\plotone{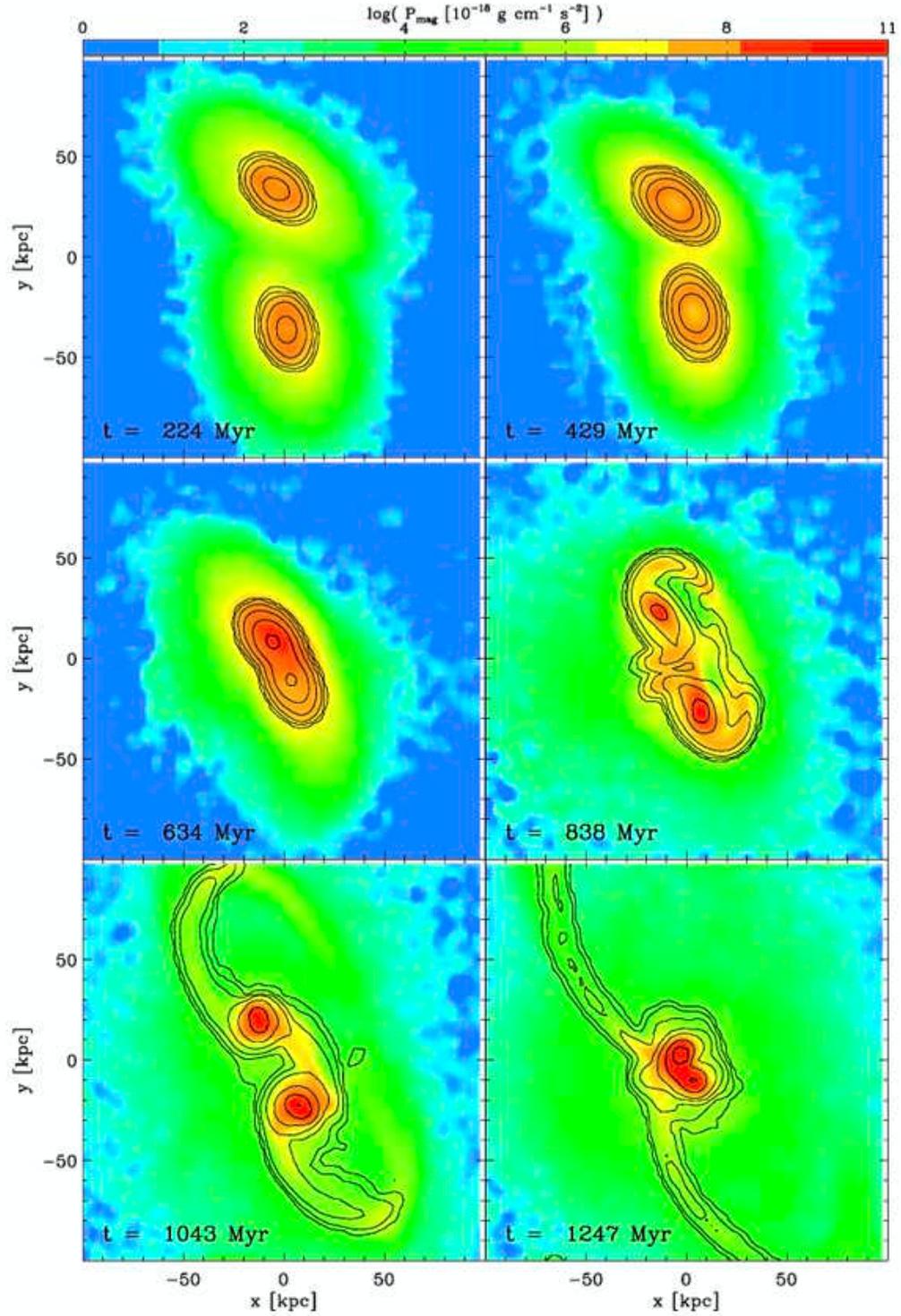}
  \caption{The Antennae simulation with an initial field of $B_0=10^{-6}$ G. Colors visualize the line-of-sight magnetic pressure $P_\mathrm{mag}=B^2/8\pi$ (in units of $10^{-18}$ g cm$^{-1}$ s$^{-2}$) and contours correspond to stellar surface density $\Sigma_\mathrm{stars}$. The contour levels are 0.005, 0.02, 0.08, 0.32, 1.28, 5.12 and 20.48 $M_\odot$ pc$^{-2}$.
  \label{evolution}}
\end{center}
\end{figure*}

Fig. \ref{evolution} shows the line-of-sight magnetic pressure $P_\mathrm{mag}= B^2/8\pi$ in the simulation with $B_0=10^{-6}$ G at different time steps, overlaid with contours of the stellar surface density $\Sigma_\mathrm{stars}$. The particle data has been transferred to a grid of 80$\times$80 cells using the TSC procedure (Triangular Shaped Cloud, see e.g. \citealp{Hockney&Eastwood1988}). As it takes roughly 600 Myr (i.e. approximately two half mass rotation periods) before the first encounter of the galaxies, the magnetic field has enough time to redistribute and form a realistic configuration in each of the galaxies prior to the merger (upper left and right panel, see also section \ref{SETUP_ISOLATED} and Fig. \ref{bfield}). The formation of the tidal arms is visible in the stellar density distribution but also in the distribution of magnetic pressure (central right and lower left panel). At time of best fit (lower right panel) most of the gas has been driven into the central region of the Antennae system. Thus, the magnetic pressure reaches its highest values in this region.

The temporal evolution of the absolute values of the magnetic fields for the simulations with different initial field values is shown in Fig. \ref{Bevol} (black dashed, black, dark blue, blue and green lines). In all cases (except for the run with $B_0=10^{-4}$ G), similarly to the simulations of the isolated galaxies (red and yellow lines), we see a mild amplification of the initial magnetic field in the beginning of the simulation due to the winding process. However, as the disks are not oriented parallel to the $xy$-plane, this initial amplification is slightly weaker than in the isolated disks. The reason is that the initial magnetic field now does not lie in the plane of the disks and thus the radial component of the magnetic field is weaker compared to the simulations of the isolated galaxy. In the case with the weakest initial field the magnetic field gets amplified by more than two orders of magnitude during the interaction, whereby the most violent amplification occurs during the first encounter at $t\approx650$ Myr. In the case with $B_0=10^{-6}$ G, however, the amplification is relatively modest. The evolution of the magnetic field for the simulation with the highest initial field (dashed line) is different: At the very beginning of the simulation, the high magnetic overpressure drives the gas out of the galaxies, thus ''blowing`` them up. Consequently, the magnetic field decreases by one order of magnitude within 100 Myrs due to attenuation and continues to decrease until the first encounter. At the time of the first encounter, it is only very weakly amplified. At the time of best match, the value of the magnetic field is approximately 10 $\mu$G within the numerical precision, independent of the initial seed field. This is roughly half the value derived from observations. The origin of this discrepancy might be observational as well as numerical and will be briefly addressed in section \ref{RADIO_APPLICATIONS}.

\subsubsection{Numerical stability}\label{NUMERICAL_STABILITY}

Fig. \ref{divB} shows the arithmetic mean of the numerical divergence $h\nabla\cdot\textbf{B}/|\textbf{B}|$ as a function of time for the Antennae simulations with the different initial magnetic field strengths (black dashed, black, dark blue, blue and green lines). For each simulation, there is an increase of the divergence during the first encounter, whereby the value of the numerical divergence increases with decreasing initial magnetic field. This is not surprising: If the magnetic tension is strong enough to overcome the gas pressure, the Lorentz force acts on the particles in a way that magnetic tension is released. On the other hand, if the magnetic pressure is significantly weaker than the gas pressure, chaotic motions of the particles driven by the encounter can fold the magnetic field on small scales - as small as the smoothing length - until the magnetic tension becomes dominant. This leads to a more irregular magnetic field and a higher numerical divergence. Thus, the numerical divergence is lowered in the presence of a stronger magnetic field.

\begin{figure*}
\begin{center}
\epsscale{0.9}
\plotone{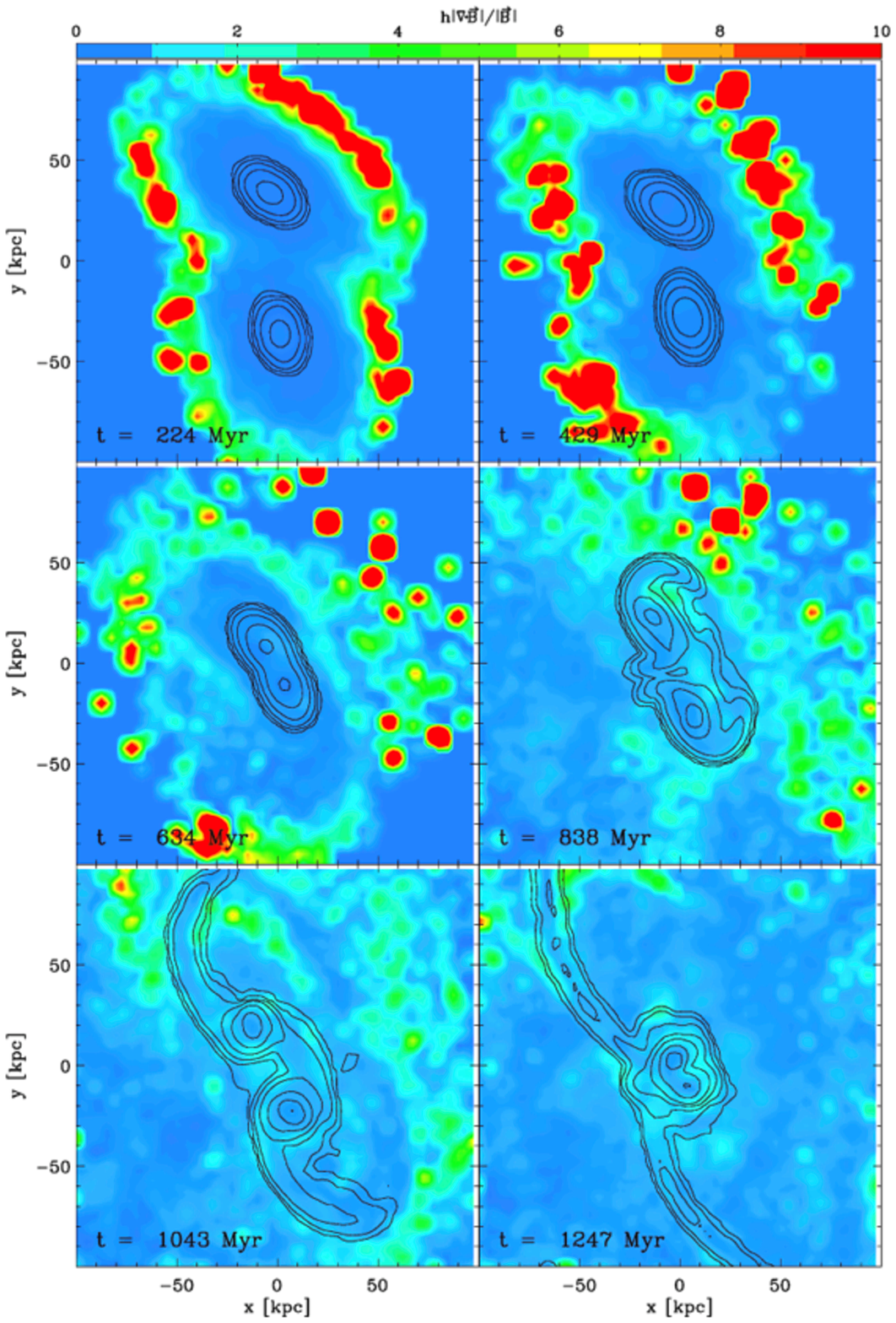}
  \caption{The Antennae simulation with an initial field of $B_0=10^{-6}$ G. Colors visualize the mean line-of-sight numerical divergence $h\nabla\cdot\textbf{B}/|\textbf{B}|$ and contours correspond to stellar surface density $\Sigma_\mathrm{stars}$. The contour levels are 0.005, 0.02, 0.08, 0.32, 1.28, 5.12 and 20.48 $M_\odot$ pc$^{-2}$.
  \label{evolution_divb}}
\end{center}
\end{figure*}

Fig. \ref{evolution_divb} shows the mean line-of-sight numerical divergence $h\nabla\cdot\textbf{B}/|\textbf{B}|$ in the simulation with $B_0=10^{-6}$ G at different time steps, overlaid with contours of the stellar surface density $\Sigma_\mathrm{stars}$ to indicate the morphology of the galaxies. The particle data has been transferred to a grid using the TSC procedure as before in Fig. \ref{evolution}. Before the first encounter (central left panel), there are regions of high (of the order of 10) numerical divergence at the ``edges'' of the galaxies (upper panels, compare also Fig. \ref{evolution}). This high numerical divergence measures can be ascribed to defective SPH calculations in these regions. The particle density there decreases to zero due to the vacuum boundary conditions (which are usually used in this type of simulations). Thus, the particle distribution within one smoothing length changes rather abruptly. Some SPH operators, including the divergence operator, are not well sampled in such a situation, leading to high numerical errors in theses estimators. As soon as the particle distribution is smoothed out as a consequence of the interaction (central left to last panel, compare also Fig. \ref{evolution}), this effect vanishes. However, it is only a small fraction of particles which are affected by this defective calculation. Thus, the arithmetic mean of the numerical divergence is lower in the beginning of the simulation than after the first encounter (Fig. \ref{divB}). Furthermore, comparing Fig. \ref{evolution} with Fig. \ref{evolution_divb} shows that in regions with the highest magnetic field values the numerical divergence is relatively low.

We have performed the same simulation with an initial magnetic field of $10^{-6}$ G but without applying the Lorentz force (not shown). In this simulation the magnetic field got amplified extremely violently by orders of magnitude to clearly unphysical values after the first encounter. The magnetic field got amplified much above the maximal value seen in the simulations presented above, and did not converge. This behaviour shows, that it is actually the Lorentz force, i.e. the backreaction of the magnetic field on the gas, which constrains the amplification. The unrealistic violence of the amplification can be traced back to the high $\nabla\cdot \textbf{B}$ values of several hundreds developing in this simulation. However, as applying the Lorentz force helps to lower the divergence in SPMHD simulations, these results are not surprising.

This can also be seen in simulations including the Lorentz force, but starting with an initially very weak magnetic field. We have performed an additional simulation with $B_0=10^{-20}$ G (not shown). In this simulation, the divergence grew to a maximal value of 2.5 during the first encounter, subsequently dropping again to values below unity. The magnetic field - and thus the Lorenz force - was very weak in this simulation, nevertheless, the divergence was still lowered to values of order unity. The magnetic field was amplified by ten orders of magnitude to a value of $10^{-10}$ G during the first encounter, which is still four orders of magnitude lower than the maximal value seen in Fig. \ref{Bevol}. This demonstrates that one can not start with an arbitrary low magnetic field and end up at micro-gauss levels after the first encounter. However, as the subsequent interaction between the two galaxies drives further turbulence, the magnetic field continued to grow after the first encounter at a rate of approximately one order of magnitude per 100 Myr. Thus, at a time of 1.1 Gyr, the magnetic field reached a value of $\approx10^{-6}$ G and slowly converged towards the maximal value seen in Fig. \ref{Bevol}. Note that during this steady growth of the magnetic field the numerical divergence was actually decreasing.

In the simulation with $B_0=10^{-4}$ G, the numerical divergence measure grows up to a value of approximately 0.4 already at the beginning of the simulation. This is because the high magnetic pressure ''blows up`` the galaxies and thus excites strong turbulent motions which in turn result in a more irregular magnetic field. We note that this behaviour shows that the value of the numerical divergence mainly depends on the irregularity of the magnetic field, which is also the reason why the numerical divergence remains relatively small in the simulations of the quiescent evolution of the isolated galaxies (red line). This can also be understood theoretically: The numerical SPH divergence-operator calculates the weighted sum of the differences of the magnetic field of a particle and its neighbouring particles within a smoothing length. Thus, the higher the degree of irregularity of the magnetic field, the higher the numerical divergence. However, the numerical divergence should not be mistaken for a physical divergence, as it is only a measure of numerical small-scale (i.e. smaller than one smoothing length) fluctuations of the field. This can be seen in simulations using Euler-Potentials, where the physical divergence is zero by definition, but the numerical divergence has still values of order unity (see \citealp{me2009}) for a tangled magnetic field. Thus, lowering the numerical divergence below this tolerance value of unity should be sufficient to guarantee a physically meaningful evolution of the magnetic field. Using the Euler-Potentials in the Antennae simulations would most probably result in a much weaker amplification of the magnetic field which should not be considered physical, because Euler-Potentials are not suitable for simulations of kinematically vigorous system (see section \ref{NUMERICS}).

\begin{figure}
\begin{center}
\epsscale{1.2}
\plotone{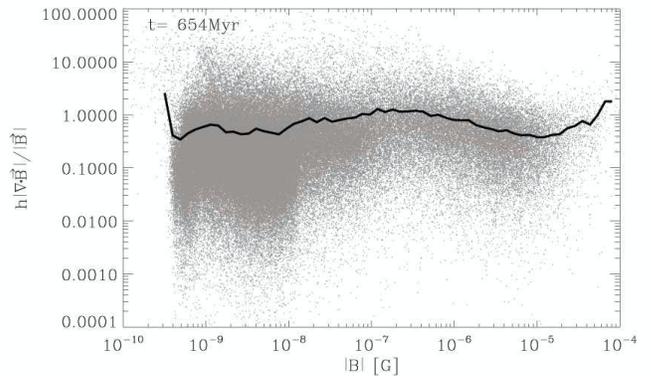}
  \caption{$h\nabla\cdot\textbf{B}/|\textbf{B}|$ as a function of the total magnetic field for the Antennae simulation with $B_0=10^{-9}$ G at time of the first encounter ($t \approx 650$ Myr). Grey dots correspond to the values of each particle, the solid line is the mean value for a given magnetic field strength. The values of the numerical divergence are widely distributed over the range of magnetic field strength.
  \label{divversb32}}
\end{center}
\end{figure}

\begin{figure}
\begin{center}
\epsscale{1.2}
\plotone{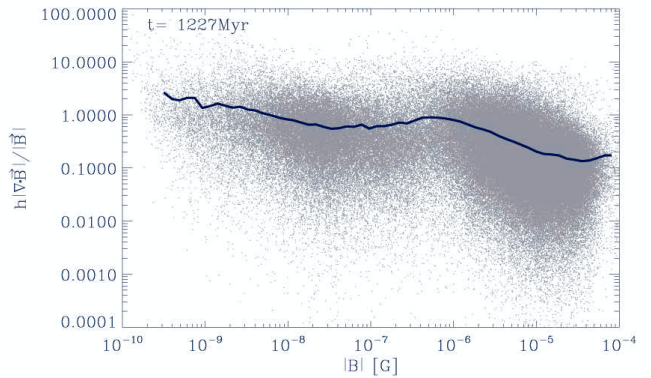}
  \caption{Same as Fig. \ref{divversb32}, but at the time of best fit ($t_\mathrm{BM} \approx 1250$ Myr). The values of the numerical divergence are widely distributed over the range of magnetic field strength and even lower for higher magnetic field values.
  \label{divversb60}}
\end{center}
\end{figure}

Moreover, the applied SPH implementation is geared to ensure that the numerical divergence measure does not alter the evolution equations for the magnetic field (see \citealp{GadgetMHD}). Thus, even if the divergence operator measures a numerical divergence, it does not influence the magnetic field evolution directly. This has been shown by \citet{Price&Monaghan2005SPMHDIII}, who demonstrated that a magnetic monopole can be advected without causing numerical instabilities. It can also be seen comparing Fig. \ref{Bevol} and Fig \ref{divB}: The lower the initial magnetic field, the weaker the magnetic field shortly after the first encounter, although the numerical divergence is higher for lower initial fields. Thus, there is no direct dependance of the magnetic field strength on the value of the numerical divergence. Fig. \ref{divversb32} and \ref{divversb60} show the numerical divergence as a function of the magnetic field strength at the time of the first encounter and at the time of the best fit, respectively. We show the plots for the simulation with $B_0=10^{-9}$ G, as the amplification of the magnetic field is the most efficient and the numerical divergence is the highest in this simulation. Thus, a possible dependance of the magnetic field on the divergence measure should be the best visible in this simulation. However, there is no significant correlation, and the values of the numerical divergence are widely distributed over the range of magnetic field strength. At the time of best fit, they are even lower for higher magnetic field values (see also Fig. \ref{evolution_divb}). This behaviour is qualitatively the same for all initial magnetic field values. Of course, the amplification is more efficient for lower initial fields, thus one could argue that it is the amplification efficiency which depends on the numerical divergence value. However, in the beginning of the simulation with $B_0=10^{-4}$ G the magnetic field actually \textit{decreases} with increasing numerical divergence showing that non-vanishing numerical divergence not necessarily leads to an amplification of the magnetic field. Rather, the amplification efficiency is restricted by the strength of the Lorentz force: The higher the magnetic field, the stronger the Lorentz force braking the motions which lead to an amplification of the magnetic field. Thus, the lower the initial field, the more efficient its amplification. Hence we conclude that as long as the numerical divergence remains as low as the numerical divergence seen in simulations with Euler-Potentials (i.e. lower than unity), the evolution of magnetic fields in SPH simulations is physically meaningful.

\subsubsection{Self-regulation of the amplification}\label{PRESSURES}

The magnetic field is expected to get enhanced through field line compression in shocks and field line stretching in shear flows. However, in the framework of MHD, any motion of gas leading to an amplification of the magnetic field will be suppressed by the magnetic field itself via the Lorentz force as soon as the magnetic energy gets comparable to the kinetic energy of the gas. The magnetic energy is then converted into kinetic energy of the gas, thus maintaining equipartition between the magnetic and gas kinetic energy density, or equivalently, the magnetic and the hydrodynamic pressure $P_\mathrm{hyd}=1/2\rho v^2$.  In particular, the magnetic field is expected to be in equipartition with the turbulent energy of the gas (see e.g. \citealp{Beck2007} and \citealp{Chyzy2007}), as only velocity gradients can lead to an amplification of the magnetic field via the induction equation. Thus, the self-regulation of the strength of the magnetic field seen in our simulations can be ascribed to equipartition between the turbulent and magnetic pressures. In order to analyze this behavior, we have examined the central region of the system in more detail, and also performed a comparison simulation without magnetic fields. We define the turbulent pressure as $P_\mathrm{turb}=1/2\rho v_\mathrm{turb}^2$, with the turbulent velocity of the i-th particle defined as

\begin{eqnarray}
  v_\mathrm{turb}(i)=\frac{1}{3}\sqrt{\sum_k{v_\mathrm{turb}^k(i)^2}},
\end{eqnarray}
 where $k=x,y,z$ and
 \begin{eqnarray}
 v_\mathrm{turb}^k(i)=\sqrt{\frac{\sum_{j=1}^N{v^k(j)-v^k(i)}}{N}},
 \end{eqnarray}
with $N=64\pm5$ being the number of the nearest neighbors.

We briefly note that the thermal pressure does not directly affect the evolution of the magnetic fields. According to the induction equation of MHD, the magnetic field evolution is determined by the velocity field alone.

\begin{figure*}
\begin{center}
\epsscale{1.1}
\plotone{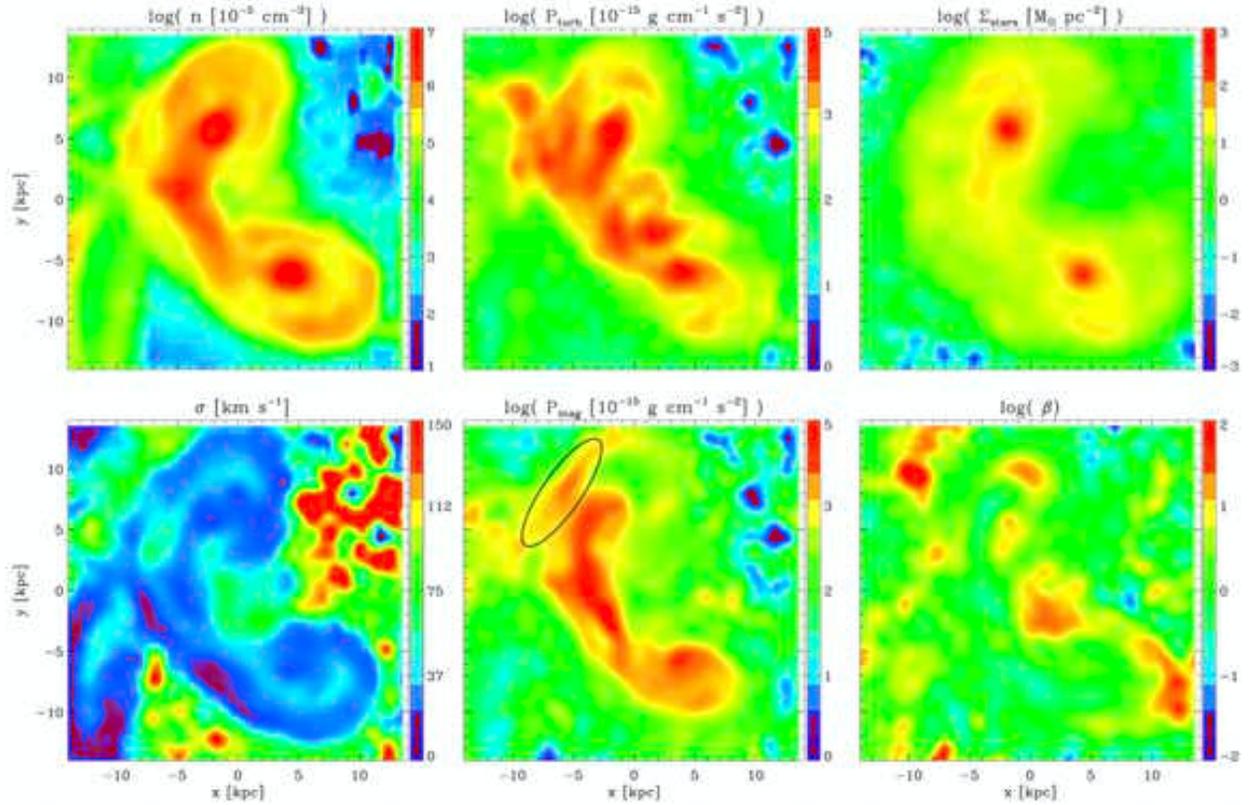}
  \caption{The Antennae simulation with an initial magnetic field of $B_0=10^{-6}$ G. From left to right and top to bottom: Mean line-of-sight gas number density $n$, turbulent pressure $P_\mathrm{turb}=1/2\rho v_\mathrm{turb}^2$, stellar surface density $\Sigma_\mathrm{stars}$, gas velocity dispersion $\sigma$, magnetic pressure $P_\mathrm{mag}=B^2/8\pi$ and $\beta=P_\mathrm{turb}/P_\mathrm{mag}$ in the inner region (innermost 28 kpc) of the system at time of best fit ($t_\mathrm{BM} \approx 1250$ Myr).
  \label{pres_mag}}
\end{center}
\end{figure*}

\begin{figure*}
\begin{center}
\epsscale{1.1}
\plotone{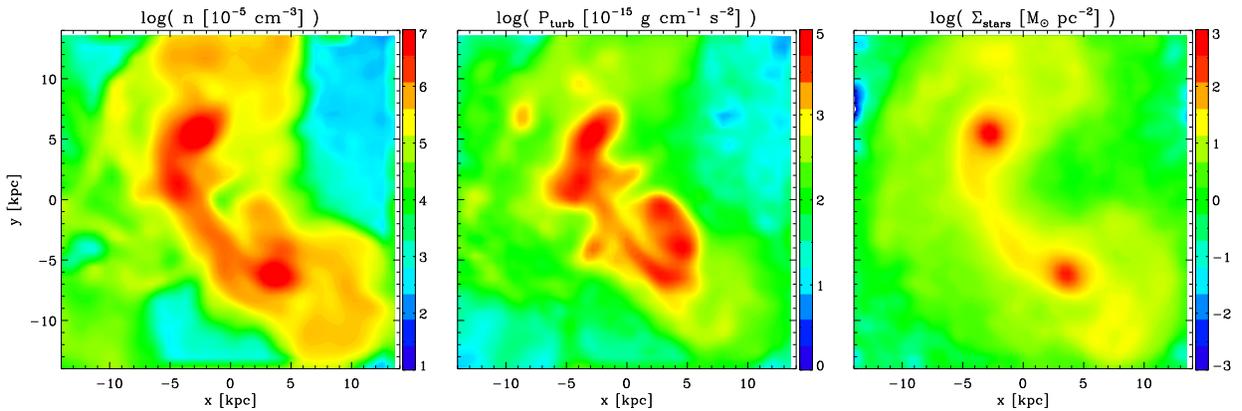}
  \caption{The Antennae simulation without including magnetic fields. From left to right: Mean line-of-sight gas number density $n$, turbulent pressure $P_\mathrm{turb}=1/2\rho v_\mathrm{turb}^2$ and stellar surface density $\Sigma_\mathrm{stars}$ in the inner region (innermost 28 kpc) of the system at time of best fit ($t_\mathrm{BM} \approx 1250$ Myr). The turbulent pressure is highest in the overlapping region between the two merging galaxies.
  \label{pres_nm}}
\end{center}
\end{figure*}

Fig. \ref{pres_mag} shows from left to right and top to bottom the gas number density $n$, the turbulent pressure $P_\mathrm{turb}$, the stellar surface density $\Sigma_\mathrm{stars}$, the velocity dispersion $\sigma$ (calculated as before in section \ref{SETUP_ISOLATED}), the magnetic pressure $P_\mathrm{mag}$ and $\beta=P_\mathrm{turb}/P_\mathrm{mag}$ in the inner region (innermost 28 kpc) of the system at time of best fit for the simulation with $B_0=10^{-6}$ G. This value of $B_0$ is comparable to the typical magnetic field value observed in spiral galaxies, which is why we have chosen this simulation for our analysis. Assuming a distance of 22 Mpc this region would comprise approximately 4.37'. \citealp{Chyzy&Beck2004} use a distance of 19.2 Mpc and observe an area of $\approx 3.5$', scaled to a distance of 22 Mpc this gives an area of approximately 22.4 kpc across. Thus, our model has a bigger extent by a factor $\approx28/22.4=1.25$, which still is in qualitative agreement with observations (see section \ref{ANT_PROP}). The particle data has been transferred to a spatial grid using the TSC procedure and averaged over the $z$-direction (i.e. the line-of-sight) with $z\in[-14\mbox{ kpc}, 14\mbox{ kpc}]$ and $z=0$ defined as the center of mass of the system. The turbulent and magnetic pressures are given in units of $10^{-15}$ g cm$^{-1}$ s$^{-2}$, corresponding to $6.242\times 10^{-4}$ eV cm$^{-3}$, i.e. the highest values are approximately 100 eV cm$^{-3}$.

In order to be able to recognize whether the magnetic field itself has a significant effect on the turbulent pressure in the system, we have applied the same analysis to a simulation without magnetic fields. Fig. \ref{pres_nm} displays from left to right the mean line-of-sight gas number density $n$, turbulent pressure $P_\mathrm{turb}$ and stellar surface density $\Sigma_\mathrm{stars}$ in the inner region (innermost 28 kpc) of the not magnetized system at time of best fit (calculated as before in Fig. \ref{pres_mag}).

Comparing Fig. \ref{pres_mag} and Fig. \ref{pres_nm} shows that in the simulation with magnetic fields (Fig. \ref{pres_mag}) the gas distribution is more compact, whereas the turbulent pressure distribution is ''disrupted``. Particulary in the northern (upper) galaxy the turbulent pressure distribution is more extended in the magnetized case (Fig. \ref{pres_mag}) than in the simulation without magnetic fields (Fig. \ref{pres_nm}). These differences probably develop because the gas is more likely to move along magnetic field lines than perpendicular to them and thus the velocity distribution is altered. The gas distribution is in both cases more extended than the stellar distribution, with the stellar density being highest in the centers of the galaxies ($\approx 10^{3}$ $M_\odot$ pc$^{-2}$). The stellar distribution is not significantly changed in the presence of a magnetic field. Since the galaxies have a low gas fraction (20\%), the total gravitational potential is dominated by the stellar component in the inner region of the Antennae system. Thus, the distribution of the high density gas ($>10$ cm$^{-3}$) is almost unaffected by the presence of the field. As stars are expected to form in high density regions, it is not surprising that the distribution of stars formed in our simulations is also independent on the presence of the field. In the magnetized case (Fig. \ref{pres_mag}), the gas velocity dispersion $\sigma$ (lower left panel) is of the order of 10 to 20 km s$^{-1}$ within the galaxies. The distribution of magnetic pressure (lower central panel in Fig. \ref{pres_mag}) is slightly different compared to the distribution of the turbulent pressure (upper central panel): The highest turbulent pressures are reached in the centers of the galaxies, whereas the magnetic pressure is highest in the overlapping region of the galaxies. Moreover, there is a ridge of magnetic pressure at the root of the southern tidal tail (indicated by the black oval) which is not visible in the distribution of turbulent pressure. This differences most probably originate in the magnetic field being a vector instead of a scalar quantity. A fully random magnetic field is not amplified efficiently by isotropic compression. Thus, only in regions with strong shear flows which stretch and therefore straighten the magnetic field it can be amplified efficiently. However, the energy range of the magnetic pressure is overall comparable to the energy range of the turbulent pressure. Thus, $\beta=P_\mathrm{turb}/P_\mathrm{mag}$ (lower right panel) is in the rage 1 to 10 almost everywhere, which means that the magnetic pressure is of the order of the turbulent pressure or slightly lower.

\begin{figure}
\begin{center}
\epsscale{1.3}
\plotone{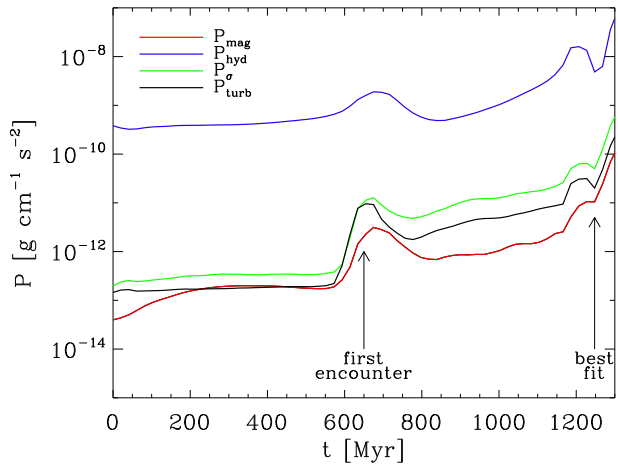}
  \caption{Temporal evolution of $P_\mathrm{turb}$ (black line), $P_\mathrm{hyd}$ (blue line), $P_\sigma$ (green line) and $P_\mathrm{mag}$ (red line) for gas particles with a number density $> 0.005$ cm$^{-3}$ in the simulation with $B_0=10^{-6}$ G. $P_\mathrm{turb}$ and $P_\mathrm{mag}$ are of a comparable order of magnitude throughout the simulation and almost in equipartition at time of best fit.
  \label{Pres_t}}
\end{center}
\end{figure}

Fig. \ref{Pres_t} shows the temporal evolution of the turbulent pressure $P_\mathrm{turb}\sim v_\mathrm{turb}^2$ (black line), the hydrodynamic pressure $P_\mathrm{hyd}\sim v^2$ (blue line) the magnetic pressure $P_\mathrm{mag}\sim B^2$ (red line) and the ''dispersion pressure``, corresponding to the velocity dispersion, i.e. $P_\sigma=1/2\rho\sigma^2$ (green line) for gas particles with a number density $> 0.005$ cm$^{-3}$ in the simulation with $B_0=10^{-6}$ G. The hydrodynamic pressure is higher than the magnetic pressure by roughly three orders of magnitude throughout the simulation, which should be expected from theory as it is not the value of the velocity itself, but the velocity gradients which determine the evolution of magnetic fields. The turbulent, dispersion and the magnetic pressures are of the same order of magnitude until the first encounter (except of the beginning of the simulation). After the encounter, the turbulent and the dispersion pressures are always slightly higher (by a factor of approximately five and ten, respectively) than the magnetic pressure. At time of best fit the turbulent and magnetic pressures are again of the same order of magnitude, as already indicated in the last panel in Fig. \ref{pres_mag}. The fact, that the magnetic pressure never exceeds the turbulent pressure indicates, that the magnetic field amplification is restricted to magnetic field values corresponding to the equipartition level between turbulent and magnetic pressure. This is exactly what is expected from theory and explains the self-regulated saturation of the magnetic field strength in our simulations (Fig. \ref{Bevol})

\begin{figure}
\begin{center}
\epsscale{1.3}
\plotone{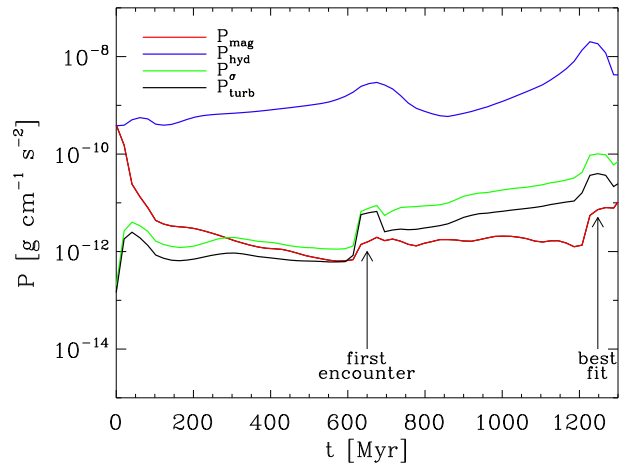}
  \caption{Same as Fig. \ref{Pres_t} but for the simulation with $B_0=10^{-4}$ G. $P_\mathrm{mag}$ is much higher than  $P_\mathrm{turb}$ in the beginning of the simulation, but decreases to the level of equipartition within 400 Myr. $P_\mathrm{turb}$ and $P_\mathrm{mag}$ are almost in equipartition at time of best fit.
  \label{Pres_t_m4}}
\end{center}
\end{figure}

Fig. \ref{Pres_t_m4} shows the same quantities as in Fig \ref{Pres_t} but for the more extreme simulation with $B_0=10^{-4}$ G. In the beginning of this simulation, the magnetic pressure is three orders of magnitude higher than the turbulent pressure (because this initial magnetic field is two orders of magnitude higher than the expected equipartition value of several $\mu$G and $P_{mag}\propto B^2$). Within the first 50 Myr of evolution the magnetic pressure drops by one order of magnitude. Simultaneously, the turbulent and dispersion pressures increase by the same amount. This is because the high magnetic pressure ''blows up`` the galaxies in the very beginning of the simulation and thus drives a lot of turbulent (or chaotic) motions. After the first 50 Myr, the difference between the turbulent and the magnetic pressure is only one order of magnitude and the system is able to relax again. Thus, the turbulent and dispersion pressures start to decrease, and the magnetic pressure continues to decrease further. After approximately 400 Myr the magnetic pressure is of the order of the turbulent and dispersion pressures. Shortly before the first encounter, it has reached a value slightly below the turbulent pressure. In the subsequent evolution, similar to the simulation with $B_0=10^{-6}$ G, the magnetic pressure always stays below the turbulent pressure. However, the evolution of the pressure components is altered compared to the simulation with $B_0=10^{-6}$ G. Particularly, the second encounter (visible as a temporary increase of the pressure values) preceding the time of best fit in Fig. \ref{Pres_t} is shifted by approximately 100 Myr to later times in the simulation with $B_0=10^{-4}$ G. This difference develops because the strong magnetic field in the beginning of the simulation with $B_0=10^{-4}$ G alters the gas distribution significantly and thus changes the evolution of the whole system. In summary, this comparison clearly shows that interacting galactic systems always tend to reach equipartition, independent of the initial ratio of magnetic to turbulent pressure.

As already discussed in section \ref{NUMERICAL_STABILITY}, without applying the Lorentz force the magnetic field gets amplified much above the value of equipartition between magnetic and turbulent pressure, and does not converge. Thus, it is actually the Lorentz force, i.e. the backreaction of the magnetic field on the gas, which yields the self-regulation.

Finally, we compared the SF rates in the simulations with different initial magnetic field strengths with the SF rate in a simulation without any magnetic field (not shown). The SFR after the first encounter in the simulation with $B_0=10^{-6}$ G showed to be slightly lower (by a factor of approximately two) than in the simulation without or with a weak magnetic field, indicating that the presence of the magnetic field hinders the collapse of gas. However, this influence is not strong enough to alter the SF history significantly.

\section{Simulated radio emission and polarization maps}\label{RADIO_MAPS}
\subsection{Computation method}\label{RADIO_METHOD}
In order to compare our results directly with observations, we compute artificial radio emission and polarization maps from our simulation data.
For this purpose, the magnetic field components and the stellar density from the SPMHD simulations have been again transferred to a three-dimensional grid. The following calculations have been performed with an IDL code developed by \citet{Wiatr}.
The calculations of the total and polarized synchrotron intensity and the calculation of the polarisation angle have been performed in the standard way according the to the following formulae (see \citealp{Longair1997Vol2} and \citealp{RybickiLightman1986} for more details):

 The total synchrotron emission $J_\nu$ at a given frequency $\nu$ is given by
 \begin{eqnarray}
   J_\nu &=&\left[\frac{4}{3}\frac{\sigma_T c^3}{\mu_0}\sqrt{\frac{\pi^3 m_e^5}{2e^3}}\left(\sqrt{\frac{2\pi m_e^3}{e}c^2}\right)^{-p}\right]\notag\\
   && \times\kappa\sqrt{B^{1+p}\nu^{1-p}},\label{Jnu}
 \end{eqnarray}
 where the magnetic field $B$ is the only input from our simulations. The frequency $\nu$ and the index of the power spectrum of the relativistic cosmic ray (CR) electrons $p$ are input parameters. The latter is assumed to be 2.6 in this paper, corresponding to the value given by \citealp{Chyzy&Beck2004}. The constants are the Thompson cross section $\sigma_T=0.665\times 10^{-24}$ cm$^2$, the magnetic permeability $\mu_0=1$ (in CGS units), the speed of light $c$ and the electron mass $m_e\approx9.1\times 10^{-28}$ g. The constant normalization factor $\kappa$ of the cosmic ray energy spectrum can be derived for a given total CR energy $E_{CR}$ via
 \begin{eqnarray}
   E_{CR}=\kappa\int_{E_{min}}^{E_{max}}E^{1-p}dE.\label{ecr}
 \end{eqnarray}
 The observed CR energies in the Milky Way follow a steep spectrum from $10^9$ to $10^{20}$ eV, whereby supernova remnants (SNR) are the most likely source for CRs with energies $<10^{18}$ eV. CRs with higher energies may be produced in Jets of pulsars or black holes, and are probably of extragalactic origin (see \citealp{Auger2007}). Given the steep fall-off of CR abundance with energy we assume an energy range of $E_{min}=10^9$ eV to $E_{max}=10^{15}$ eV in our calculations. Furthermore, as CRs in this energy range are most likely produced in SNRs, we assume the CR distribution to be proportional to the stellar density, with a typical value of the mean specific energy density of $e_{CR}=1$ MeV m$^{-3}$ for CR protons (see e.g. \citealp{Ferriere2001}). However, we apply a cutoff at an energy density of $e_{CR}=100$ MeV m$^{-3}$. The energy density of CR electrons is roughly 100 times lower than the energy density of CR protons, thus the mean energy density for CR electrons is assumed to be $10$ keV m$^{-3}$.

 $J_\nu$ is calculated within every grid cell at a frequency of $\nu=4.86\times10^9$ Hz (corresponding to the observed frequency). The total intensity $I_\mathrm{tot}$ of the synchrotron radiation is subsequently obtained by integration of the emission along the line-of-sight.

The degree of polarization $\Pi$ of any electromagnetic radiation is defined as the amount of its polarized intensity $I_p$ compared to the amount of its total intensity $I_\mathrm{tot}$. The synchrotron emission of a single radiating charge is always polarized elliptically, because the light component for which polarization is parallel to the magnetic field projected onto the plane of sky ($I_\|$) has a different refraction index than the perpendicular component ($I_\bot$). However, as charges gyrate along the magnetic field lines, the elliptical components will cancel, as emission cones will contribute equally from both sides of the line-of-sight. Thus, for any reasonable distribution of particles that varies smoothly with pitch angle, the radiation will be partially linearly polarized and thus characterized by the terms $I_\|$ and $I_\bot$. The degree of linear polarization for particles of a single energy can then be expressed as
\begin{eqnarray}
  \Pi(\nu)=\frac{I_\bot(\nu)-I_\|(\nu)}{I_\bot(\nu)+I_\|(\nu)},\label{PI}
\end{eqnarray}
where $I_{tot}(\nu)=I_\bot(\nu)+I_\|(\nu)$. If the energy spectrum of the radiating particles follows a power-law (here $N(E)=\kappa E^{-p}dE$), it can be shown that
\begin{eqnarray}
  \Pi=\frac{p+1}{p+\frac{7}{3}}.\label{PI2}
\end{eqnarray}
Thus, in a homogeneous magnetic field, the degree of polarization is very high (approximately 73\% for $p=2.6$). However, when integrated along the line-of-sight, opposite polarization cancels out and the observed degree of polarization is therefore usually much lower than the theoretically expected value.

The polarized intensity $I_p$ depends on the Stokes parameters $Q$ and $U$ according to
\begin{eqnarray}
  I_p=\sqrt{Q^2+U^2},
\end{eqnarray}
with
\begin{eqnarray}
  Q&=&\Pi \int J_\nu\cos{(2\psi)} ds,\\
  U&=&\Pi \int J_\nu\sin{(2\psi)} ds,
\end{eqnarray}
where the integration is performed along the line-of-sight and $\psi$ is the polarization angle, defined as the angle between the electric field vector of the radiation perpendicular to the magnetic field ($\vec{E}_\bot$) and the $x$-axis in the $xy$-plane (the plane of sky), i.e.:
\begin{eqnarray}
  \sin{(2\psi)}&=&-\frac{2B_xB_y}{B_x^2+B_y^2},\\
  \cos{(2\psi)}&=&\frac{B_x^2-B_y^2}{B_x^2+B_y^2}.
\end{eqnarray}

Finally, the observed degree of polarization is
\begin{eqnarray}
  \Pi_\mathrm{obs}=\frac{I_p}{I_\mathrm{tot}}.
\end{eqnarray}

All calculated values are subsequently convolved with a telescope beam corresponding to the 17''$\times$14'' beam in the observations of \citet{Chyzy&Beck2004}, i.e. the beam-diameter is approximately 1.5 kpc at the distance of the Antennae system (assuming a distance of 22 Mpc). The shape and sensitivity of the beam is specified by a 2D gaussian function.

\subsection{Applications}\label{RADIO_APPLICATIONS}
\begin{figure}
\begin{center}
\epsscale{1.2}
\plotone{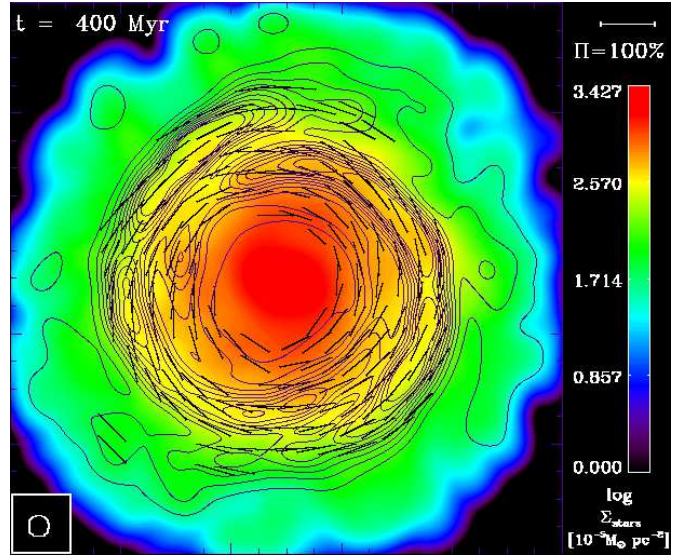}
  \caption{Face on view of the isolated galaxy. Colors correspond to the logarithm of stellar surface density (in units of $10^{-5}$ $M_\odot$ pc$^{-2}$), overlaid with contours of total synchrotron power. The contour levels are 0.001 to 0.01 mJy in ten equally spaced steps. Magnetic field lines derived from calculations of polarization are shown in black.
  \label{RadioGalaxy}}
\end{center}
\end{figure}

Fig. \ref{RadioGalaxy} shows an simulated face-on radio map of the isolated disk at $t=400$ Myr for the simulation with $B_0=10^{-6}$ G. The particle data of a domain with $x\in[-20 \mbox{ kpc},20 \mbox{ kpc}]$, $y\in[-20\mbox{ kpc},20\mbox{ kpc}]$ and $z\in[-10\mbox{ kpc},10\mbox{ kpc}]$ (with the zero-point defined as the center of mass of the system) has been transferred to a spatial grid with $60\times60\times30$ cells. Thus, the displayed domain comprises $40\times40$ kpc$^2$. The colours correspond to the stellar surface density, overlaid with contours of total synchrotron power. Magnetic field lines derived from calculations of polarization are shown in black. To account for the spatial isotropy of the emission from any emitting volume element, the total flux has been multiplied by the factor

 \begin{eqnarray}
   f_\mathrm{obs}=\frac{\pi \cdot r_\mathrm{beam}^2}{4\pi\cdot d^2},
 \end{eqnarray}

 with $d=22$ Mpc the distance to the observer and $r_\mathrm{beam}$ the assumed radius of the beam.
 Thus, the artificial flux, given in mJy, corresponds to what is expected to reach the earth from the distance of the Antennae system.

As already discussed above, the initially homogeneous magnetic field gets redistributed by the differential rotation of the disk, thus developing a spiral pattern which is clearly visible in the total emission. The magnetic field lines trace this spiral pattern. Altogether, the structure of the magnetic field is similar to what is observed in typical disk galaxies. A similar result has been also obtained independently by \citet{Kulesza2009}, who have performed 3D MHD simulations of barred spiral galaxies using a grid code.

Interestingly, the distribution of the magnetic field lines derived from the polarization calculations (Fig. \ref{RadioGalaxy}) does not extend as far out in the disk as the magnetic field itself (Fig. \ref{bfield}). This difference occurs because we can only observe polarization where enough CR particles are present, which is not the case in the outer parts of the galaxy. However, the structure of the magnetic field is comparable.

\begin{figure}
\begin{center}
\epsscale{1.2}
\plotone{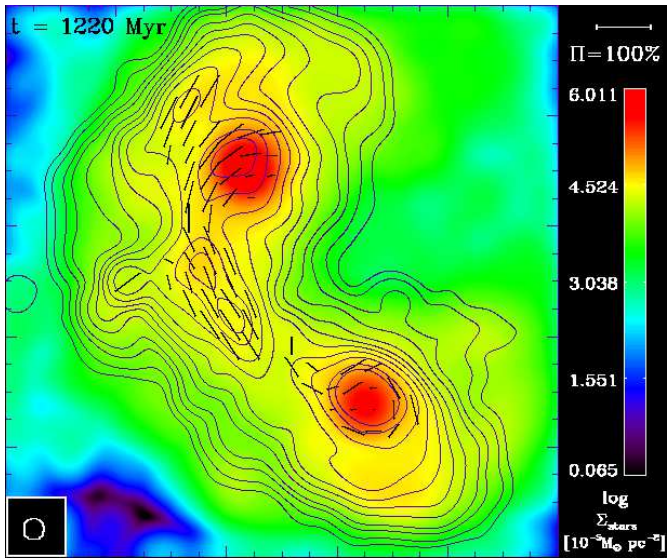}
  \caption{Inner region (innermost 28 kpc) of the simulated Antennae system. Colors correspond to the stellar surface density (in units of 10$^{-5}$ $M_\odot$ pc$^{-2}$), overlaid with contours of total synchrotron power. The contour levels are 0.005, 0.12, 0.30, 0.53, 1.2, 2.1, 3.3, 5.3, 9.0, 17 and 24 mJy. Magnetic field lines derived from calculations of polarization are shown in black. The simulated systems compares very well to the observed system (Fig. \ref{Antennae2}).
  \label{RadioAntennae}}
\end{center}
\end{figure}

Fig. \ref{RadioAntennae} shows a simulated radio map of the inner region of the Antennae system for the simulation with $B_0=10^{-6}$ G at the time of best match ($t_\mathrm{BM} \approx 1.25$ Gyr). The particle data of a domain with $x\in[-14 \mbox{ kpc},14 \mbox{ kpc}]$, $y\in[-14\mbox{ kpc},14\mbox{ kpc}]$ and $z\in[-56\mbox{ kpc},56\mbox{ kpc}]$ (with the zero-point again defined as the center of mass of the system) has been transferred to a spatial grid with $75\times75\times300$ cells. The total flux was again corrected to the isotropy of the emission by multiplying by the factor $f_\mathrm{obs}$. The contour levels of total synchrotron emission have been chosen to be as given in Fig. 3 in \citet{Chyzy&Beck2004}. Thus, they are the same levels as displayed in Fig. \ref{Antennae2}.

Given the fact that our simulations are fully self-consistent, the similarity between the simulated and the observed system is astonishing. The spatial extent and distribution of the total synchrotron flux compares very favorably with the observations. Also, the highest values of total synchrotron emission are reached in the overlapping region and at the centers of the interacting galaxies. Furthermore, two ridges of ordered magnetic field lines, one reaching from one galaxy to the other along the overlap region, and one corresponding to the root of the southern tidal tail, naturally develop in our simulation. However, there are also several differences: There is a lack of magnetic fields in the southern tidal tail, i.e. the ordered magnetic field structure is not as prominent as in the observations, which may be caused by a lack of CRs in this region. Also, there is no western spiral arm in NGC4038 (the upper galaxy), which is probably because the spiral structure of the progenitors in our simulation is not pronounced enough. Furthermore, there is too little polarized emission in the outskirts of the galaxies and the overlap region is shifted north (down) compared to the observations. Moreover, the pitch angle of the magnetic field in the isolated galaxy (Fig. \ref{RadioGalaxy}) is rather small. The latter can probably be explained by the absence of a dynamo process in this simulation. Despite these differences the satisfactory match between observation and our simulation is encouraging. Thus, our numerical method already seems to capture the most essential basic processes relevant in investigating interactions of magnetized galaxies. In particular, our model of the Antennae system seems to provide a fair description of how this system may have formed.

A further discrepancy between observation and our simulation is the value of the magnetic field itself. The magnetic field strength in our simulations saturates at a mean value of roughly 10 $\mu$G, only 10\% of the simulated particles carry magnetic field values of $|B|>20\mu$G and only 1\% have $|B|>50\mu$G. On the other hand, the mean magnetic field strength derived from observations of synchrotron radiation is approximately 20 $\mu$G. However, the observed value is derived assuming equipartition between the CR energy and the energy of the magnetic field, an assumption which does not necessarily have to hold. Furthermore, the assumed CR energy density and the magnetic dissipation factor in our calculations are only approximate estimates. Given these uncertainties, a difference by a factor of two between the observed magnetic field and the field strength in our simulations is admissible.

\section{Conclusion and Discussion}\label{SUMMARY}
We have presented the first fully self-consistent \textsl{N}-body/SPH simulations of the interacting Antennae galaxy system including magnetic fields. We show that weak magnetic seed fields in the isolated disk galaxies are amplified by the gravitational interaction throughout the two galactic encounters. Thereby the magnetic pressure saturates at a level corresponding to equipartition between the turbulent and the magnetic pressure, independent of the initial field strength. Particularly, magnetic fields with an initial value higher than the equipartition value diminish during the evolution, demonstrating that the state of equipartition is the natural state for magnetized galactic systems. An analysis of artificial total synchrotron emission and polarization maps provides a convincing agreement with the observations. Summarizing, the method of \textsl{N}-body/SPH simulations including magnetic fields reproduces quite conclusively the complicated dynamics of the amplification and spatial design of magnetic fields in interacting galaxies.

Moreover, a detailed discussion of the numerical divergence of the magnetic field in SPH simulations has been presented in section \ref{NUMERICAL_STABILITY}. Our analysis strongly suggests that numerical divergence measures which are smaller than a certain threshold can be considered as measures of sub-resolution fluctuations which do not affect the overall evolution of the magnetic field. Considering simulations using the Euler-Potentials, which pose a $\nabla\cdot\textbf{B}$-free prescription by definition, this threshold can be assessed to be $h\nabla\cdot\textbf{B}/|\textbf{B}| \approx 1$ (see also \citealp{me2009}).

What can we learn from these simulations for the global evolution of cosmic magnetic fields? Within the framework of standard CDM hierarchical clustering models the formation of large disk galaxies as well as elliptical galaxies is characterized by more or less intense merging of smaller galactic subunits, e.g. dwarf galaxies, collapsing gas clouds or globular clusters. If we assume that at least some of the accreted subunits have been magnetized by stellar activity (e.g. supernova explosions, stellar winds or T-Tauri-jets), the merging of such subunits to larger galaxies must have been accompanied by a significant amplification and restructuring of the magnetic field on galactic scales. The amplification and ordering of small-scale magnetic fields to a toroidal configuration during the evolution of isolated galaxies was recently shown by \citet{Hanasz2009} and \citet{Dubois&Teyssier2009} independently. \citet{Hanasz2009} considered an axially symmetric galactic disk in which stellar seed fields were amplified by a cosmic ray driven dynamo. \citet{Dubois&Teyssier2009} demonstrated the amplification and ordering of small-scale fields seeded by SF activity in the context of the formation of a dwarf galaxy with significant galactic winds. Complementary to these findings, our simulations prove that amplification via non-axisymmetric three dimensional gravitational interaction alone may provide an alternative channel for galactic as well as intergalactic magnetic field evolution. In other words, given that the structure formation is characterized by a galactic bottom-up architecture, we would expect that within one or two Giga-years the Universe has been globally magnetized by the combination of dynamo action in isolated galaxies and dynamical amplification by interacting galactic objects. However, dynamo action is supposed not to be very efficient in dwarf galaxies since their differential rotation is not strong enough (\citealp{Gressel2008}). Thus, at an early epoch of the universe, when most of the galaxy population consists of dwarfs, magnetic field amplification due to interactions may be even more significant.

With their study of the formation of dwarf galaxies including magnetic fields and galactic winds, \citet{Dubois&Teyssier2009} demonstrate an alternative scenario based on the ideas of \citet{Bertone2006} which they call the "Cosmic Dynamo". According to their findings, galactic winds from young dwarf galaxies eject magnetic field energy into the intergalactic medium, leading to a mean intergalactic field $B_\mathrm{IGM}$ of $10^{-11}$ to $10^{-10}$ G.  The preceding amplification of the magnetic field inside the dwarf galaxy by the combined action of stellar activity and differential rotation (i.e. the Galactic Dynamo) is thereby restricted by the IGM magnetic field already present at the formation time of the galaxy. For an IGM magnetic field of $B_\mathrm{IGM}\approx 10^{-10}$ G, the Lorentz force may prevent the formation of a new generation of dwarf galaxies and subsequent star formation. As a consequence, the IGM magnetic field never grows significantly above $10^{-10}$ G. Since dwarf galaxies are characteristic in the early phase of the evolution of the universe, this Cosmic Dynamo may have been very efficient in magnetizing the IGM. However, besides the accretion of IGM material previously enriched with magnetic fields, \citet{Dubois&Teyssier2009} also point out the importance of accretion of satellite galaxies for the evolution and amplification of the magnetic field in galaxies at later times.

Our simulations emphasize the Cosmic Dynamo scenario proposed by \citet{Dubois&Teyssier2009}. The efficient amplification of the magnetic field during the equal-mass-merger presented here clearly shows that interactions of galaxies should be taken into account in studies of the magnetic evolution of the universe. We would also expect an intergalactic medium which is not only enriched with heavy elements by stellar activity, but also magnetized on large scales by galaxy interactions. Our simulations may help to understand the observationally well established facts that very young galaxies already exhibit magnetic field strengths comparable with nearby fully developed spiral galaxies and the rotation measure estimates of intergalactic magnetic fields (e.g. \citealp{Bernet2008}).

\begin{acknowledgements}

We thank the referee for very helpful comments which improved the paper significantly. H. Kotarba is grateful for interesting discussions with Krzysztof Chy{\.z}y and Michal Hanasz. The authors also thank Krzysztof Chy{\.z}y for providing Fig. \ref{Antennae2}. This research was supported by the DFG Cluster of Excellence ``Origin and
Structure of the Universe'' (\url{www.universe-cluster.de)}. S. Karl and T. Naab acknowledge support from the DFG priority program SPP1177.

\end{acknowledgements}

\nocite{*}
\bibliographystyle{apj}
\bibliography{Antennae}

\label{lastpage}

\end{document}